\documentclass[twocolumn,showpacs,superscriptaddress,prb]{revtex4}
\usepackage{makeidx}
\usepackage{graphicx}
\usepackage{dcolumn}
\usepackage{bm}
\usepackage{amsmath}
\usepackage{amsfonts}
\usepackage{amssymb}

\begin{document}

\title{Cooling a Micro-mechanical Beam by Coupling it to a Transmission Line}
\author{Fei Xue}
\affiliation{CREST, Japan Science and Technology Agency (JST), Kawaguchi, Saitama
332-0012, Japan}
\affiliation{Frontier Research System, The Institute of Physical and Chemical Research
(RIKEN), Wako-shi, Saitama 351-0198, Japan}
\author{Y.D. Wang}
\affiliation{NTT Basic Research Laboratories, NTT Corporation, Atsugi-shi, Kanagawa
243-0198, Japan}
\author{Yu-xi Liu}
\affiliation{CREST, Japan Science and Technology Agency (JST), Kawaguchi, Saitama
332-0012, Japan}
\affiliation{Frontier Research System, The Institute of Physical and Chemical Research
(RIKEN), Wako-shi, Saitama 351-0198, Japan}
\author{Franco Nori}
\affiliation{CREST, Japan Science and Technology Agency (JST),
Kawaguchi, Saitama 332-0012, Japan}
\affiliation{Frontier Research
System, The Institute of Physical and Chemical Research (RIKEN),
Wako-shi, Saitama 351-0198, Japan}
\affiliation{Center for
Theoretical Physics, Physics Department, Applied Physics Program,
Center for the Study of Complex Systems, The University of Michigan,
Ann Arbor, Michigan 48109-1040, USA}


\date{\today}

\begin{abstract}
We study a method to cool down the vibration mode of a
micro-mechanical beam using a capacitively-coupled superconducting
transmission line. The Coulomb force between the transmission line
and the beam is determined by the driving microwave on the
transmission line and the displacement of the beam. When the
frequency of the driving microwave is smaller than that of the
transmission line resonator, the Coulomb force can oppose the
velocity of the beam. Thus, the beam can be cooled. This mechanism,
which may enable to prepare the beam in its quantum ground state of
vibration, is feasible under current experimental conditions.
\end{abstract}


\maketitle

\section{Introduction}

Mechanical resonators~\cite{Clelandbook2002,Blencowe2004} have
important applications in high precision displacement
detection~\cite{Caves1980,Bocko1996,LaHaye2004}, mass
detection~\cite{Buks2006pre}, quantum
measurements~\cite{Braginskybook1992}, and studies of quantum
behavior of either mechanical
motion~\cite{Mancini2002,Marshall2003,Eisert2004,
Wei2006,Xue2007prb,Xue2007NJP} or
phonons~\cite{Hu1996a,Hu1996b,Hu1997,Hu1999}. Recently,
proposals~\cite{Savel'ev2004Dec,Savel'ev2006NJP, Savel'ev2007} have
been made for implementing qubits by using buckling nanoscale bars
with quantized motion. Casimir effects on nanoscale mechanical
device were also studied~\cite
{Chan2001,Munday2005,Munday2006,Capasso2007}. However, in previous
studies (see, e.g.,
Refs.~\onlinecite{Savel'ev2004Dec,Savel'ev2006NJP, Savel'ev2007}) of
quantized mechanical resonators (and macroscopic quantum
phenomena~\cite{Leggett2002} in mechanical resonators, see, e.g.,
Refs.~\onlinecite{Mancini2002,Marshall2003,Eisert2004}), it is
necessary to prepare the mechanical resonators into their
vibrational ground states. Therefore, one needs to cool the
mechanical resonators down to ultra-low temperatures to put them
into their ground states. For example, a temperature below one
milli-Kevin is necessary for cooling a 20 MHz mechanical resonator
to its vibrational ground state.

To reach temperatures below one milli-Kelvin, which is beyond the
capability of present dilution refrigerators, alternative cooling
mechanisms are now being explored. Using optomechanical couplings,
the cooling of mechanical resonators was recently demonstrated
experimentally~\cite
{Metzger2004,Arcizet2006,Gigan2006,Kleckner2006,Schliesser2006,Poggio2007}.
To observe the quantized motion of a mechanical resonator, one
should be able to cool the mechanical resonator down to its ground
state of vibration and to detect the phonon number state. Besides
optomechanical cooling, electronic
cooling~\cite{Martin2004Mar,Zhang2005,Wineland2006,Naik2006,Wang2007,Zhao2007}
was also studied. For instance, theoretical proposals for cooling a
mechanical resonator were considered by coupling it either to a
two-level system~\cite{Martin2004Mar,Zhang2005,Wang2007}, to an
ion~\cite{Hensinger2005} or to an LC circuit~\cite{Wineland2006}. An
experimental demonstration of cooling a mechanical resonator by the
quantum back-action of a superconducting single-electron transistor
was recently reported~\cite{Naik2006}. Most of these cooling
experiments (e.g.,
Refs.~\onlinecite{Metzger2004,Arcizet2006,Gigan2006,Kleckner2006})
focus on cooling mechanical resonators with a frequency lower than 1
MHz, with a mechanical quality factor higher than $10^{4}$. It is
difficult to experimentally cool mechanical resonators to their
quantum ground state of vibrations because of the weak coupling
between the mechanical resonators and the cooling media for
optomechanical systems (see, e.g., Ref.~\onlinecite{Bernad2006}).

Recently, the strong coupling between a one dimensional (1D)
transmission line resonator (TLR) and a solid state
qubit~\cite{You2003prb,You2005PT} was achieved~\cite{Wallraff2004},
and the detection of photon number states was also
demonstrated~\cite{Schuster2007}. Based on these experimental
developments, here we consider replacing the Fabry-P\'erot cavity
used in previous cooling proposals~\cite{Metzger2004} by a 1D TLR,
in order to cool a micron-scale bar.

The working mechanism of our proposal here is similar to the cooling
of a tiny mirror in a Fabry-P\'erot cavity~\cite{Metzger2004}. This
cooling mechanism can be summarized as follows. A force on the
mirror is coupled to the light intensity inside the cavity. This
intensity does not change instantaneously with each mirror
displacement. The delayed response of the intensity to a change in
the mirror displacement leads to a force that can either agree or
oppose the motion of the mirror, depending on whether the laser
frequency is bigger or smaller than the cavity resonant
frequency~\cite{Milonni2004}. By including this intensity-dependent
force, in addition to a thermal force on the mirror, the mirror can
be cooled.

In our proposal here, the TLR, whose frequency is determined by its
overall capacitance and inductance, acts as a cavity. The beam is
placed near the middle of the TLR and capacitively coupled to the
TLR. When the mechanical beam has a displacement, the overall
capacitance of the TLR changes, thereby the resonant frequency of
the TLR also changes. Now let us consider the case where the TLR is
driven by a microwave with fixed frequency. Any displacement of the
beam will change, after a delay, the voltage between the TLR and the
beam (and also the force between them). Recall that here we are
considering two coupled oscillators: the TLR and the mechanical
beam. The rf microwave drive acts directly on the TLR, and
indirectly on the mechanical beam. After the transients are gone,
the driven damped oscillator (here, the TLR) exhibits a steady-state
response which is delayed with respect to the drive. In other words,
the beam displacement changes the TLR's oscillation frequency
$\omega_a$. Since the frequency $\omega_{\rm rf}$ of the drive is
fixed, this change in $\omega_a$ will affect the steady-state
amplitude of the TLR oscillator, which will be reached after some
delay. The displacement of the beam (i.e., the action on the TLR),
causes a delayed reaction (i.e., a delayed back-action) force from
the TLR to the beam. The delay is determined by the damping rate of
the TLR. When the frequency of the microwave $\omega_{\rm rf}$ is
smaller than the resonant frequency $\omega_{a0}$ of the TLR, this
back-action force opposes the motion the beam, thereby damping the
Brownian motion of the beam.

This cooling mechanism studied here is also related to the mechanism
recently employed in Refs.~\onlinecite{Brown2007,Wineland2006}.
There, cooling is produced by a capacitive force which is
phase-shifted relative to the cantilever motion. In their set-up,
when the cantilever oscillates, its motion modulates the capacitance
of an LC circuit, therefore modulating its resonant frequency. This
resonant frequency, and the potential across the capacitance, is
modulated relative to the fixed frequency of the applied rf drive.
The modulated force linked to this potential shifts the resonant
frequency of the cantilever~\cite{Brown2007,Wineland2006}. Because
of the finite response time of the LC circuit, there is a phase lag
in the force, relative to the motion. When the rf frequency is
smaller than the resonant frequency, the phase lag produces a force
that opposes the cantilever velocity, producing damping. When this
damping is realized without introducing too much noise in the force,
then the cantilever is cooled.

Our analysis, presented below, shows that it is possible to cool a 2
MHz beam, initially at $\sim50$ mK, down to its quantum vibrational
ground state at around $0.07$ mK. This is a cooling factor of about
$1/700$. Our proposed device, which is a combination of the devices
in Refs.~\onlinecite{Wallraff2004,Naik2006}, should be realizable in
experiments. Moreover, because of its on-chip structure, our device
has some practical advantages to be integrated in dilution
refrigerators and be operated on; while optomechanical systems need
an additional optical system.

\begin{figure}[bp]
\centering
\includegraphics[bb=140 490 465 760, clip, width=8.5cm]{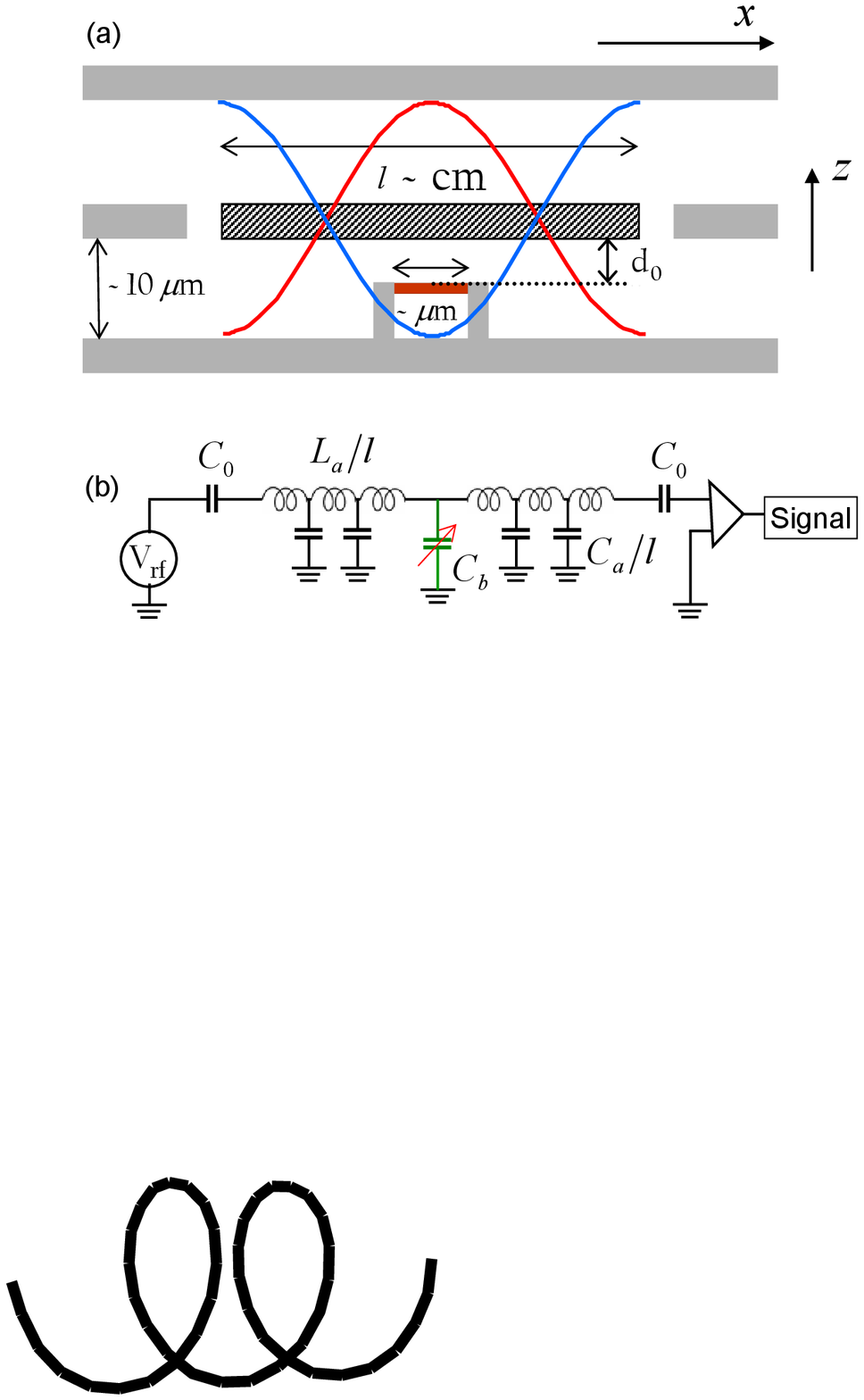}
\caption{(Color online) (a) Schematic layout for the proposed device
and (b) its equivalent circuit. A superconducting coplanar
stripline, forming a 1D TLR, provides a cavity. A doubly-clamped
beam (central small dark-red rectangle) is placed between two
horizonal superconducting lines. This red beam is capacitively
coupled to the central (hatched) superconducting line at a maximum
of the voltage standing-wave in the 1D TLR. The capacitances $C_{0}$
allow the input and output signals to be coupled to the central
(hatched) stripline. This allows to measure the amplitude and phase
of the 1D TLR and applying dc and rf pulses to the 1D TLR.}
\label{fig:device1}
\end{figure}

\section{Device}

Our proposed device is illustrated in Fig.~\ref{fig:device1}(a). A
doubly-clamped micro-beam is placed in the middle of a 1D
superconducting TLR formed by thin coplanar striplines. The central
stripline has a length $l$, with a capacitance $C_{a}/l$ and an
inductance $L_{a}/l$, per unit length. For not-very-high
frequencies, the equivalent circuit of the stripline is an infinite
series of inductors with each node capacitively connected to the
ground, as shown in Fig.~\ref{fig:device1}(b). It can be described
as a series of resonators that accommodate different resonant
modes~\cite{Wallraff2004}. Since the length of the micro-beam is
much smaller than that of the 1D TLR, we consider the voltage in the
middle of the 1D TLR to be the voltage $V_{a}\left( t\right) $ on
the beam. Here, we only consider the mode with the largest coupling,
i.e., the lowest mode~\cite{Wallraff2004} coupled to the beam. The
1D TLR is coupled to both, two semi-infinite TLRs to the left and
right, via the capacitors $C_{0}$, and the beam via the capacitor
$C_{b}$. Thus the boundary conditions and the voltage of the 1D TLR
are modified by these additional capacitors. When $C_{0}\, , C_{b}
\ll C_{a}\,$, the circuit can be approximated by a 1D TLR with a
modified frequency
\begin{equation}
\omega _{a}= \frac{1}{\sqrt{L_{a}C_{t}}} \, {\rm ,}
\end{equation}
with $C_{t}=C_{a}+C_{b}+2C_{0}$. Actually, due to its coupling to
the environment, the 1D superconducting TLR acts as a cavity with
finite quality factor $Q_{a}$ $=$ $\omega _{a}/2\gamma $, where
$2\gamma$ is the damping rate of the 1D TLR.

\begin{figure}[tp]
\centering
\includegraphics[bb=165 550 430 640, clip, width=8.5cm]{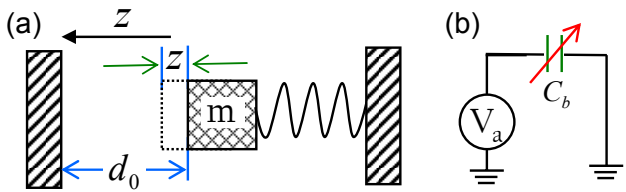}
\caption{(Color online) (a) Schematics of the beam (mass and spring)
capacitively coupled to a conductor and (b) its equivalent circuit.
The charging energy of the beam is determined by the voltage $V_{a}$
and variable capacitor $C_{b}$ between the beam and the 1D
transmission line. } \label{fig:device2}
\end{figure}

The fundamental vibration mode of the doubly-clamped beam can be
approximated by a mechanical resonator with frequency $\omega _{b}$
and effective mass $m$. The beam is coupled to a conductor (the 1D
TLR) via a capacitor, and its equivalent circuit is illustrated in
Fig.~\ref{fig:device2}. The beam is exposed to a Coulomb force from
the 1D TLR. Please note that for the case we studied in this paper,
the amplitudes of the oscillates are small and thus the beam is
essentially in the linear regime. For a review of nonlinear
oscillators, see, e.g., Ref.~\onlinecite{Dykman1984}.

\section{Coulomb force on the beam}

\begin{figure}[tp]
\centering
\includegraphics[bb=200 550 400 650, clip, width=8.5cm]{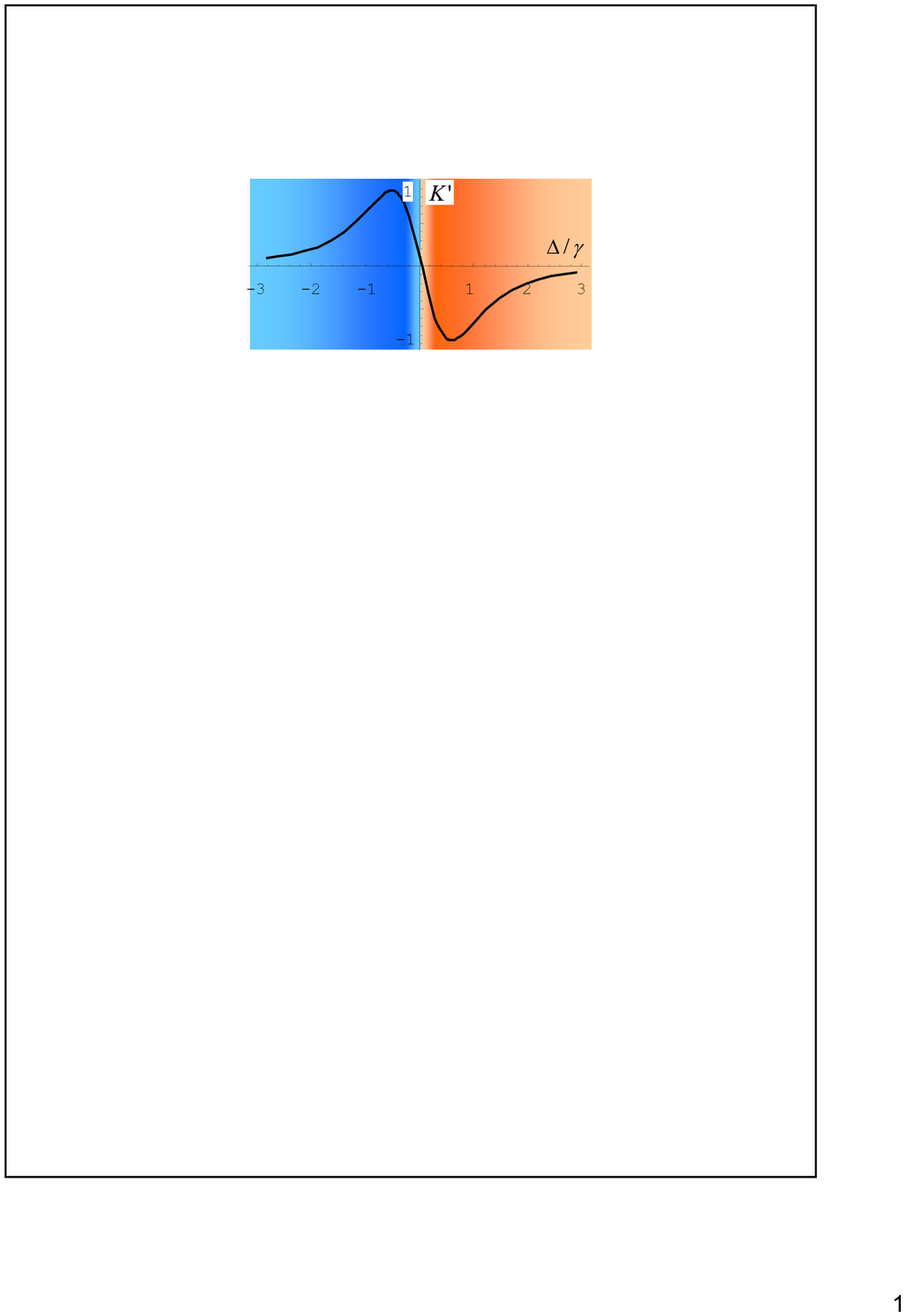}
\caption{(Color online) The effective elastic constant $K^{\prime}$
(arbitrary units) versus the detuning $\Delta = \omega _{\text{rf}}
- \omega _{\text{a0}}$ (scaled by $\gamma$). When $\omega
_{\text{rf}} < \omega _{\text{a0}}$, the beam can be cooled. Here,
the region $\Delta>0$~($\Delta<0$) is shaded by the red~(blue)
color, representing the heating~(cooling) of the beam,
respectively.} \label{fig:Kc-omega-rf}
\end{figure}

This force gives rise to a cooling mechanism which is similar to the
cavity-cooling of the vibrating mirror in
Ref.~\onlinecite{Metzger2004}. As shown in
Fig.~\ref{fig:device2}(a), it is assumed that the beam vibrates
around its equilibrium position with an amplitude $z(t) \equiv z$,
which is much smaller than the distance $d_{0}$ between its
equilibrium position and the TLR, i.e., $z \ll d_{0}$. The averaged
Coulomb force on the beam can be written as~\cite{Wineland2006}
\begin{equation} \label{eq:force1}
F_{C}\left( z\right) =\frac{d_{0}}{4\left( d_{0}-z\right) ^{2}}%
\,C_{\rm b0}V_{a}^{2}\left( z\right) \text{,}
\end{equation}%
when $\omega _{a} \gg \omega _{b}$. Here, $C_{\rm b0}$ is the
capacitance between the beam and TLR for $z=0$. Assuming that an
external driving source $V_{d}=V_{\rm rf}\,\cos \left( \omega _{\rm
rf}t\right) $ acts on the central TLR via the capacitor $C_{0}$,
$V_{a}^{2}\left( z\right) $ will reach a steady amplitude after a
time delay $\tau _{d} \sim 1/\gamma $. To first order in $z$,
Eq.~(\ref{eq:force1}) can be rewritten as
\begin{equation}
F_{C}\left( z\right) \cong F_{0}+K^{\prime }\,z\text{,}
\label{eq:electrical-force}
\end{equation}%
where the effective elastic constant  $K^{\prime }$ of the Coulomb
force on the beam by the TLR is
\begin{equation}
K^{\prime }=K_{E}\,D\left( \alpha \right) \left[ 1-\alpha ^{2}\frac{C_{\rm b0}}{%
C_{\rm t0}}D\left( \alpha \right) \left( \alpha ^{2}-1+\frac{1}{Q_{a}^{2}}%
\right) \right] \text{.} \label{eq:eff-spring-constant01}
\end{equation}%
The term $F_{0}$, which is independent of the displacement of the
beam, will change the equilibrium position of the beam. $F_{0}$ does
not contribute to the cooling of the beam, and can be canceled by
applying an appropriate dc voltage between the TLR and the beam.
Therefore, hereafter it will be omitted. The term
\begin{equation}
K_{E}=\frac{C_{\rm b0}V_{\rm rf}^{2}}{2d_{0}^{2}}
\end{equation}%
describes the coupling strength between the beam and the TLR.
$C_{\rm t0}$ is the total capacitance of the TLR for $z=0$. And
\begin{equation}
D\left( \alpha \right) =\left[ \left( \alpha ^{2}-1\right) ^{2} +
\frac{\alpha ^{2}}{Q_{a}^{2}} \right] ^{-1}
\end{equation}
is a dimensionless parameter determined by the ratio
\begin{equation}
\alpha =\frac{\omega _{\rm rf}}{\omega _{a}} \text{.}
\end{equation}
$D\left( \alpha \right) $ takes its maximum value on resonance
$\omega _{\text{rf}}/\omega _{a}=1$. The typical behavior of
$K^{\prime}$, versus the detuning
\begin{equation}
\Delta =\omega _{\rm rf}-\omega _{\text{a0}} \text{,}
\end{equation}
is plotted in Fig.~\ref{fig:Kc-omega-rf}. Here, $\omega
_{\text{a0}}$ is the frequency of the TLR for $z=0$. There is an
optimal detuning point for the driving microwave where $K^{\prime}$
takes its maximum value. As shown in Fig.~\ref{fig:Kc-omega-rf}, the
sign of the effective elastic constant $K^{\prime }$ of the Coulomb
force is determined by the detuning between the frequency $\omega
_{\text{rf}}$ of the driving microwave and that of the TLR $\omega
_{\text{a0}}$. When $\omega _{\text{rf}} < \omega _{\text{a0}}$,
additional damping is induced by the Coulomb force, cooling the beam
because of its delayed response to the displacement of the beam.

\section{Cooling mechanism}

We define the effective temperature $T_{\rm eff}$ of the fundamental
vibration mode of the beam according to the equipartition law
\begin{equation}
T_{\rm eff}=K_{\text{eff}} \frac{\left\langle z^{2}
\right\rangle}{k_{B}} \text{,} \label{eq:defination-of-temperature}
\end{equation}
where $k_B$ is the Boltzmann constant, $K_{\text{eff}}$ a modified
elastic constant of the beam after considering the existence of the
TLR. In some papers (e.g. in
Refs.~\onlinecite{Metzger2004,Poggio2007}) the effective temperature
is defined/estimated from the \textit{original} elastic constant
$K_0$ of the beam, instead of the \textit{effective} elastic
constant $K_{\text{eff}}$ of the beam. For the case when the
effective elastic constant $K^{\prime}$ of the force $F_C$ is much
smaller than the original elastic constant $K_0$ of the beam, the
definition in Eq.~(\ref{eq:defination-of-temperature}) and the one
in Refs.~\onlinecite{Metzger2004,Poggio2007} give almost the same
result of the effective temperature. However, please note that for a
large effective elastic constant $K^{\prime}$ of the force $F_C$,
one should not neglect the modification of the elastic constant of
the beam.

The Coulomb force from the TLR has two effects on the mean kinetic
energy of the beam. First, because of its delayed response to the
displacement of the beam, the Coulomb force introduces additional
damping in the beam motion, thereby increases or reduces the mean
kinetic energy of the beam. Below, it is shown that the damping rate
of the beam increases when $\omega_{\text{rf}} <
\omega_{\text{a0}}$. Second, fluctuations in the Coulomb force from
the TLR also introduce additional noise in the motion of the beam,
thereby increasing the mean kinetic energy of the beam. The balance
of these two competing effects gives the theoretical lower limit of
the attainable effective temperature by this cooling mechanism.

To evaluate the cooling effect of the Coulomb force from TLR, we use
the following equation of motion for the beam~\cite{Metzger2004}:
\begin{equation}
m\frac{d^{2}z}{dt^{2}}+m\Gamma \frac{dz}{dt}+K_{0}\,z=F_{\rm th}+\int_{0}^{t}\frac{%
dF_{C}\left[ z\left( t^{\prime }\right) \right] }{dt^{\prime
}}h\left( t-t^{\prime }\right) dt^{\prime }\text{,}
\label{eq:motion-equation-NAMR}
\end{equation}%
where $\Gamma =\omega _{b}/Q_{b}$ describes the coupling strength
between the beam and its thermal environment. Here, $Q_{b}$ is the
quality factor of the beam; $K_{0}=m\omega _{b}^{2}$ the elastic
constant of the beam; $F_{\rm th}$ the thermal noise force on the
beam, with a spectral density~\cite{Sidles1995}
\begin{equation}
S_{\rm th}=4k_{B}T_{0}m\Gamma \,\text{.}
\label{eq:thermal-noise-beam}
\end{equation}%
$T_{0}$ is the temperature of the environment. $F_{C}\left( z
\right)$ is the Coulomb force on the beam, acting on the beam via a
delay-function:
\begin{equation}
h\left( t\right) =1-e^{-\gamma t} \text{,}
\end{equation}%
for $t>0$. Using the Laplace transform, we obtain the mean-squared
motion of the beam:
\begin{eqnarray}
&& \left\langle z^{2} \right\rangle = \notag \\
&& \frac{k_{B}T_{0}\, \omega _{\rm {a0}}^{2} \Gamma }{\pi
K_{0}}\int_{-\infty }^{+\infty }d\omega \left[ \left(\omega ^{2} -
 \frac{K_{{\rm
eff}}}{K_{0}}\,\omega _{a0}^{2} \right) ^{2} + \omega
^{2}\, \Gamma_{\rm{eff}}^{2}\right] ^{-1} . \notag \\
\end{eqnarray}
There are three measurable effects on the vibration mode $\omega
_{b}$ of the beam from $F_{C}\left( z \right)$: a modified effective
elastic constant $K_{\rm eff}$:
\begin{equation}
K_{\rm eff}=K_{0}\left( 1-\frac{1}{1+\beta ^{2}}\frac{K^{\prime
}}{K_{0}}\right) \text{;} \label{eq:eff-spring-constant}
\end{equation}%
a modified damping rate $\Gamma _{\rm eff}$:
\begin{equation}
\Gamma _{\rm eff}=\Gamma \left( 1+Q_{b}\frac{\beta }{1+\beta ^{2}}\frac{%
K^{\prime }}{K_{0}}\right) \text{,}  \label{eq:eff-damping-beam}
\end{equation}%
with $\beta =\omega _{b}/\gamma $; and additional noise in the
motion of the beam generated by the fluctuation of $F_{C}\left( z
\right)$.

For the case $\vert K^{\prime} \vert /K_{0} \ll 1$, neglecting
fluctuations of $F_{C}\left( z \right)$ in
Eq.~(\ref{eq:motion-equation-NAMR}),  the steady value of the
mean-squared motion of the beam is given by
\begin{equation}
\left\langle z^{2}\right\rangle \approx k_{B} T_{0}
\frac{\Gamma}{\Gamma_{\rm eff}}\text{,}
\label{eq:effective-temperature03}
\end{equation}
which defines an effective temperature of the beam. For the case
when $\vert K^{\prime} \vert \approx K_{0}$ or $\vert K^{\prime}
\vert
> K_{0}$, the frequency of the beam will be greatly shifted away
from the original one, resulting in weak cooling of the
beam~\cite{Wang2007}.

\section{Effects of fluctuations of $F_C(z)$}
In the above discussions, we did not consider the effect of
fluctuations of $F_{C}\left( z \right)$ on the effective
temperature. Equation (\ref{eq:eff-damping-beam}) shows that the
damping rate of the beam is modified because of the existence of the
TLR. Thus, the effective temperature is changed, (see
Eqs.~(\ref{eq:defination-of-temperature}) and
(\ref{eq:effective-temperature03})), however, according to the
fluctuation and dissipation theorem, dampings are always accompanied
with noises. We now study noises on the beam introduced by the force
$F_{C}$ of the TLR. Actually, There are several noise sources
affecting $F_{C}\left( z \right)$, such as fluctuations in the
driving microwave, back-action due to measurements on the TLR, and
thermal noise in the TLR. Among these noise sources, the thermal
noise provides an intrinsic limit of the fluctuations of
$F_{C}\left( z \right)$. Therefore, a lower limit $S_{\rm TLR}$ of
the spectral density of $F_{C}\left( z \right)$ can be obtained by
considering the voltage fluctuation $S_{V}$ from the thermal noise
in the TLR, which is given by $S_{V}=4k_{B}T_{0}R$, with $R=2\gamma
L_a$ the effective resistor in the TLR. The voltage fluctuation
$S_{V}$ gives rise to a fluctuation of the charge on the capacitor
$C_{b}$, giving rise to fluctuations of $F_{C}\left( z \right)$ on
the beam through the capacitor $C_{b}$. Since $C_{b} \approx C_{\rm
b0} $ for small vibration amplitudes of the beam, we find that the
thermal noise in the TLR gives a fluctuation of $F_{C}\left( z
\right)$ on the beam
\begin{equation}
S_{\rm TLR}=2 k_{B}T_{0} \, D(\alpha) R C_{\rm b0}\, K_{\rm E} \,
\text{.} \label{eq:noise-beam-TLR}
\end{equation}

Assuming an Ohmic friction for the beam, the temperature
$T^{\prime}$ and the damping rate $\Gamma ^{\prime}$ of the beam and
the spectral density $S$ of the noise on the beam have the following
relation\cite{Weissbook2001}
\begin{eqnarray}
S \propto \Gamma ^{\prime}\, \omega\, \coth\frac{\hbar \omega}{2k_B
T^{\prime}} \, .
\end{eqnarray}
For not very low temperature near the beam's resonant frequency we
find that an effective temperature of the beam could be related to
the spectral density $S$ and the damping rate of the beam:
$T^{\prime} \propto S/\Gamma ^{\prime}$.
Therefore, after considering the fluctuation and dissipation theorem
we further modify the effective temperature of the beam to
\begin{equation}
T_{\rm eff}=T_{0} \left( \frac{\Gamma }{\Gamma _{\rm eff}} \right) \frac{S_{\rm TLR}+S_{\rm th}}{S_{\rm th}}\text{%
,}  \label{eq:eff-temperature02}
\end{equation}
where we only take into account the thermal noise in the TLR.
Indeed, the attainable lowest effective temperature of the beam
would be higher than the above limit, since there are also other
fluctuations acting on the beam, from both the driving microwave and
the back-action of the measurement on the beam. These fluctuations
would add more noises, which depend on the special parameters of the
circuit for the driving microwave and the circuit for the
measurement, e.g., the noise from amplifiers~\cite{Brown2007}, in
the numerator of Eq.~(\ref{eq:eff-temperature02}). We do not address
them here.

\section{Cooling ability}

\begin{figure}[tp]
\centering
\includegraphics[bb=200 445 415 635, clip, width=8.5cm]{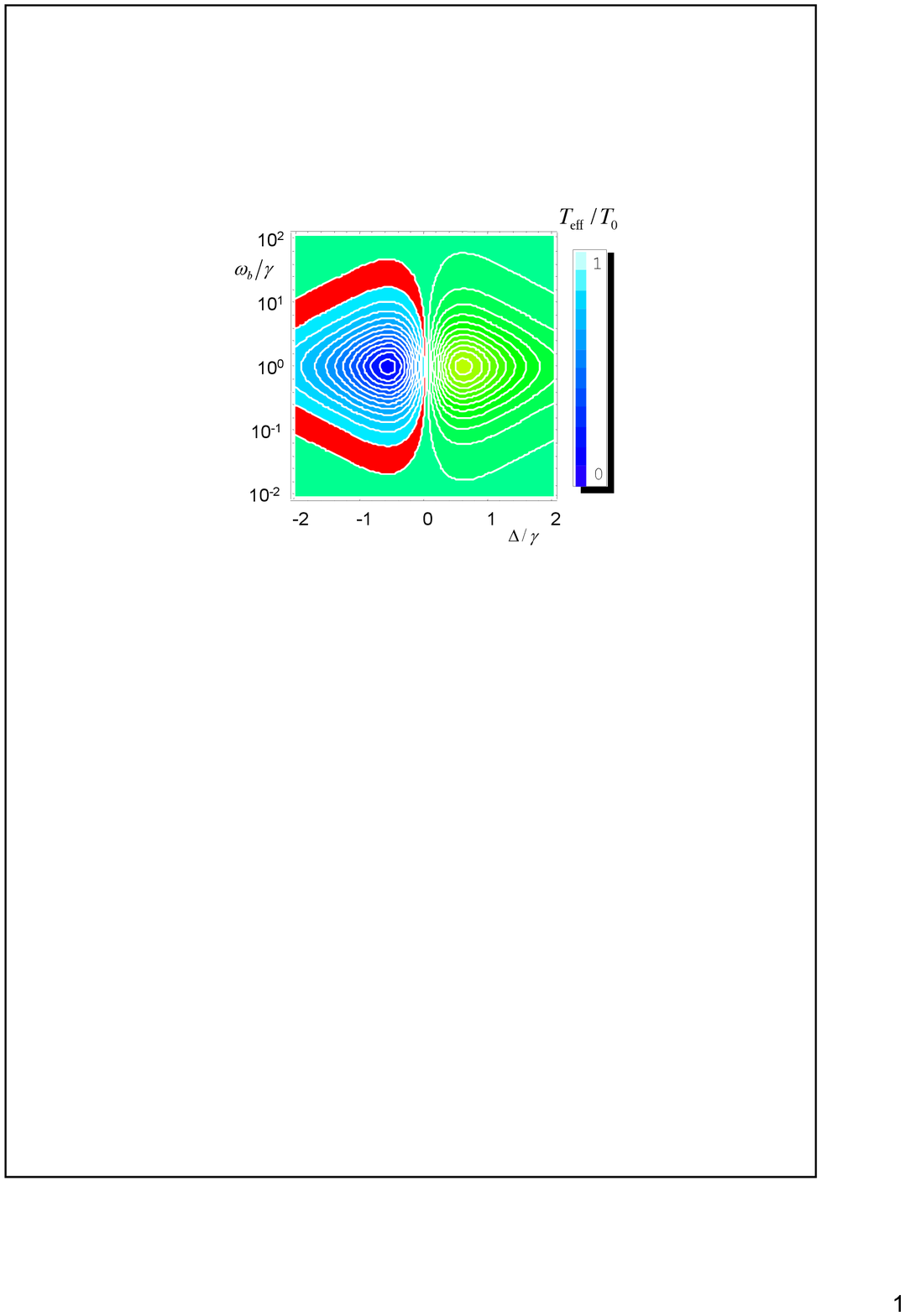}
\caption{(Color online) Cooling effect $T_{\rm eff}/T_{0}$ versus
the oscillating frequency of the beam $\omega_{b}$ and the detuning
$\Delta = \omega _{\text{rf}} - \omega _{\text{a0}}$. Both rescaled
by $\gamma$, for the typical parameters listed in the text. There is
a cooling window for $\omega_{b}$ and $\Delta$. The cooling effect
is normalized to unity. The vertical bar at right refers to cooling
effects (in blue) and not heating (in red). The darker blue
corresponds to a greater cooling effect $T_{\rm eff}/T_{0}$. The
beam is heated in the red region, and becomes unstable in the green
region.} \label{fig:efficiency}
\end{figure}

Now let us estimate the cooling effect
$T_{\rm eff}/T_{0}$. Using experimentally feasible parameters~\cite%
{Wallraff2004,Naik2006}, we take $\omega _{\rm a0}=10$ GHz,
$\gamma=200$ KHz, $d_{0}=0.1$ $\mu$m, $C_{\rm b0}=400$ aF,
$C_{\text{t0}}=10^{3}$ $C_{\rm b0}$, and $K_{0}=10$ N/m. When the
driving power of the microwave is set at $V_{\rm rf}=0.05$ mV, the
effective spring constant $K^{\prime}$ of the Coulomb force can be
as large as 0.55 N/m for optimal detuning of the driving microwave.
It is possible to obtain stronger coupling between the beam and the
TLR by increasing the driving power of the microwave, as long as the
voltage $V_{a}$ between the beam and the TLR is kept below the
breakdown voltage. Using the parameters list above and assuming
$Q_b=10^{5}$, the cooling effect $T_{\rm eff}/T_{0}$ depends on the
oscillating frequency $\omega_{b}$ of the beam and the detuning
$\Delta$, as shown in Fig.~\ref{fig:efficiency}. In
Fig.~\ref{fig:efficiency}, we assume $\vert K^{\prime} \vert \ll
K_{0}$, and then take the effective spring constant $K_{\rm eff}
\approx K_{0}$ in Eq.~(\ref{eq:eff-spring-constant}). Otherwise, the
optimal value of $\omega_{b}/\gamma$ to reach the lowest $T_{\rm
eff}$ will be slightly drifting away from unity~\cite{Wang2007}. The
best cooling effect on a 200 KHz beam is estimated to be
$T_{\text{eff}}/T_{0} \approx 3.6 \times 10^{-4}$ for the parameters
given above. Therefore, if this beam is precooled by the dilution
refrigerator to a temperature of 1 K, it can be further cooled down
to 0.36 mK using the TLR.

For a 2 MHz beam, we use a stronger microwave $V_{\rm rf}=0.5$ mV.
The best cooling effect is about $T_{\text{eff}}/T_{0} \approx 1.4
\times 10^{-3}$ with $\gamma=2$ MHz. If the beam is precooled by the
dilution refrigerator to a temperature of 50 mK, it could be further
cooled down to 0.07 mK by the TLR. This implies that the thermal
phonons in the 2 MHz beam will be less than 0.24, where a quantum
description is expected~\cite{Marquardt2007,Wilson-Rae2007}. It
should be noticed that, when the beam reaches a quantum regime a
quantum theory is expected to give the cooling efficiency in the
quantum regime.

Above we do not consider matching the impedance of the TLR to that
of the conventional microwave components~\cite{Wallraff2004}. To
obtain an optimal impedance, e.g. 50 $\Omega$, of TLR, one needs to
carefully design the geometry of the TLR and the beam. If one simply
considers the TLR as a straight coplanar transmission line, then
$C_{t0}$ might be $\sim$ 1 pF for a 10 GHz TLR. Thus, a larger
$C_{b0}$ is necessary for this larger $C_{t0}$ to maintain the
cooling effects described above.

\section{Discussions}

The fabrication of superconducting TLR now is good enough to provide
a 1D TLR with an electrical quality factor as high as $10^{6}$. The
reduced effective temperature $T_{\rm eff}$ of the beam can be
inferred from the power spectrum of the 1D TLR around the
oscillating frequency $\omega _{b}$, whose integral is proportional
to the effective temperature. Detection of microwave photon was
achieved in recent experiments, capable of resolving a single
microwave photon number~\cite{Schuster2007}. Therefore, in principle
the information of the beam could be inferred by detecting the field
in the TLR.

The working principle of our proposal is similar to that in the
optomechanical cooling by a Fabry-P\'erot cavity~\cite{Metzger2004}.
The cooling or heating is determined by the detuning between the
driving laser and cavity, which is determined by the detuning
between the driving microwave and the TLR in our case. In both
cases, a high mechanical quality factor is needed, since it measures
heating effects on the beam by its thermal environments.

Our proposal is also similar to the one in
Ref.~\onlinecite{Wineland2006}, which deals with a cantilever and a
coupled LC circuit. Here we consider a doubly-clamped beam coupled
to a co-planar transmission line resonator (TLR). Besides
considering different physical systems, in
Ref.~\onlinecite{Wineland2006}, they analyze only \textit{two}
special detunings between the frequency of the driven microwave and
the resonant frequency of the LC circuit. Here we present a more
general result for the effective elastic constant versus the
detuning, valid for all values of the detuning between the frequency
of the driven microwave and the resonant frequency of the TLR. We
also explain how the damping rate of the beam and the noise on the
beam is changed by the TLR. Our studies enable us to optimize the
setup of experimental parameters for achieving a lower effective
temperature of the beam, as shown in Fig.~\ref{fig:efficiency}.

Since the best cooling is obtained when $\omega _{b}/\gamma \approx
1$, the cooling efficiency of optical-cavity cooling would be
efficient for beams with tens of MHz, or even higher frequency,
considering current experimental parameters. For a typical optical
cavity, with a resonant frequency of $\sim 10^{14}$ Hz, the damping
rate is about $10^{8}$ Hz, for an optical quality factor $10^{6}$,
making it favorable for cooling a 100 MHz beam. However, to cool a
beam with a $\sim 1$ MHz vibration frequency, the optimal damping
rate of the optical cavity would also be $\sim 1$ MHz. This requires
an optical quality factor $\sim 10^{8}$ for a tiny mirror. A high
mechanical quality factor is also required at the same time, which
is a great challenge for the fabrication of the tiny mirror.
However, in our case, the damping rate of the TLR can be as small as
200 KHz. The damping rate of the TLR can be easily increased to
match the frequency of the beam by attaching an additional circuit
to the TLR, while it is very difficult to decrease the damping rate
of an optical cavity to match the lower-frequency beam. Thus a
MHz-beam could be cooled down to its quantum ground state and also
reach the regime $\gamma \ll \omega _{b}$, where the cavity line
width is much smaller than the mechanical frequency and the
corresponding cavity detuning. Then the photon sidebands could be
resolved when the beam is cooled down to the quantum
regime~\cite{Marquardt2007}. A recent interesting study of the lower
limit for resonator-based side-band cooling can be found in
Ref.~\cite{Grajcar2007}.

\section*{Acknowledgement}
We thank M. Grajcar, S. Ashhab and K. Maruyama for helpful
discussions. FN was supported in part by the US National Security
Agency (NSA), Army Research Office (ARO), Laboratory of Physical
Sciences (LPS), the National Science Foundation grant No.
EIA-0130383, and the JSPS CTC program. YDW was partially supported
by the JSPS KAKENHI (No.18201018).


\begin{thebibliography}{37}
\expandafter\ifx\csname
natexlab\endcsname\relax\def\natexlab#1{#1}\fi
\expandafter\ifx\csname bibnamefont\endcsname\relax
  \def\bibnamefont#1{#1}\fi
\expandafter\ifx\csname bibfnamefont\endcsname\relax
  \def\bibfnamefont#1{#1}\fi
\expandafter\ifx\csname citenamefont\endcsname\relax
  \def\citenamefont#1{#1}\fi
\expandafter\ifx\csname url\endcsname\relax
  \def\url#1{\texttt{#1}}\fi
\expandafter\ifx\csname urlprefix\endcsname\relax\def\urlprefix{URL
}\fi \providecommand{\bibinfo}[2]{#2}
\providecommand{\eprint}[2][]{\url{#2}}

\bibitem[{\citenamefont{Cleland}(2002)}]{Clelandbook2002}
\bibinfo{author}{\bibfnamefont{A.~N.} \bibnamefont{Cleland}},
  \emph{\bibinfo{title}{Foundations of nanomechanics: From Solid-State Theory
  to Device Applications}} (\bibinfo{publisher}{Springer-Verlag},
  \bibinfo{address}{Berlin}, \bibinfo{year}{2002}).

\bibitem[{\citenamefont{Blencowe}(2004)}]{Blencowe2004}
\bibinfo{author}{\bibfnamefont{M.}~\bibnamefont{Blencowe}},
  \bibinfo{journal}{Phys. Rep.} \textbf{\bibinfo{volume}{395}},
  \bibinfo{pages}{159} (\bibinfo{year}{2004}).

\bibitem{Caves1980}
\bibinfo{author}{\bibfnamefont{C.~M.} \bibnamefont{Caves}},
\bibinfo{author}{\bibfnamefont{K.~S.} \bibnamefont{Thorne}},
\bibinfo{author}{\bibfnamefont{R.~W.~P.} \bibnamefont{Drever}},
\bibinfo{author}{\bibfnamefont{V.~D.} \bibnamefont{Sandberg}}, \bibnamefont{and}
  \bibinfo{author}{\bibfnamefont{M.} \bibnamefont{Zimmermann}},
  \bibinfo{journal}{Rev. Mod. Phys.} \textbf{\bibinfo{volume}{52}},
  \bibinfo{pages}{341} (\bibinfo{year}{1980}).

\bibitem{Bocko1996}
\bibinfo{author}{\bibfnamefont{M.~F.} \bibnamefont{Bocko}} \bibnamefont{and}
  \bibinfo{author}{\bibfnamefont{R.} \bibnamefont{Onofrio}},
  \bibinfo{journal}{Rev. Mod. Phys.} \textbf{\bibinfo{volume}{68}},
  \bibinfo{pages}{755} (\bibinfo{year}{1996}).

\bibitem[{\citenamefont{LaHaye et~al.}(2004)\citenamefont{LaHaye, Buu,
  Camarota, and Schwab}}]{LaHaye2004}
\bibinfo{author}{\bibfnamefont{M.~D.} \bibnamefont{LaHaye}},
  \bibinfo{author}{\bibfnamefont{O.}~\bibnamefont{Buu}},
  \bibinfo{author}{\bibfnamefont{B.}~\bibnamefont{Camarota}}, \bibnamefont{and}
  \bibinfo{author}{\bibfnamefont{K.~C.} \bibnamefont{Schwab}},
  \bibinfo{journal}{Science} \textbf{\bibinfo{volume}{304}},
  \bibinfo{pages}{74} (\bibinfo{year}{2004}).

\bibitem[{\citenamefont{Buks and Yurke}(2006)}]{Buks2006pre}
\bibinfo{author}{\bibfnamefont{E.}~\bibnamefont{Buks}} \bibnamefont{and}
  \bibinfo{author}{\bibfnamefont{B.}~\bibnamefont{Yurke}},
  \bibinfo{journal}{Phys. Rev. E} \textbf{\bibinfo{volume}{74}},
  \bibinfo{pages}{046619} (\bibinfo{year}{2006}).

\bibitem{Braginskybook1992}
\bibinfo{author}{\bibfnamefont{V.~B.} \bibnamefont{Braginsky}},
  \emph{\bibinfo{title}{Quantum Measurement}} (\bibinfo{publisher}{Cambridge
University Press},
  \bibinfo{address}{Cambridge}, \bibinfo{year}{1992}).

\bibitem[{\citenamefont{Mancini et~al.}(2002)\citenamefont{Mancini,
  Giovannetti, Vitali, and Tombesi}}]{Mancini2002}
\bibinfo{author}{\bibfnamefont{S.}~\bibnamefont{Mancini}},
  \bibinfo{author}{\bibfnamefont{V.}~\bibnamefont{Giovannetti}},
  \bibinfo{author}{\bibfnamefont{D.}~\bibnamefont{Vitali}}, \bibnamefont{and}
  \bibinfo{author}{\bibfnamefont{P.}~\bibnamefont{Tombesi}},
  \bibinfo{journal}{Phys. Rev. Lett.} \textbf{\bibinfo{volume}{88}},
  \bibinfo{pages}{120401} (\bibinfo{year}{2002}).

\bibitem[{\citenamefont{Marshall et~al.}(2003)\citenamefont{Marshall, Simon,
  Penrose, and Bouwmeester}}]{Marshall2003}
\bibinfo{author}{\bibfnamefont{W.}~\bibnamefont{Marshall}},
  \bibinfo{author}{\bibfnamefont{C.}~\bibnamefont{Simon}},
  \bibinfo{author}{\bibfnamefont{R.}~\bibnamefont{Penrose}}, \bibnamefont{and}
  \bibinfo{author}{\bibfnamefont{D.}~\bibnamefont{Bouwmeester}},
  \bibinfo{journal}{Phys. Rev. Lett.} \textbf{\bibinfo{volume}{91}},
  \bibinfo{pages}{130401} (\bibinfo{year}{2003}).

\bibitem[{\citenamefont{Eisert et~al.}(2004)\citenamefont{Eisert, Plenio, Bose,
  and Hartley}}]{Eisert2004}
\bibinfo{author}{\bibfnamefont{J.}~\bibnamefont{Eisert}},
  \bibinfo{author}{\bibfnamefont{M.~B.} \bibnamefont{Plenio}},
  \bibinfo{author}{\bibfnamefont{S.}~\bibnamefont{Bose}}, \bibnamefont{and}
  \bibinfo{author}{\bibfnamefont{J.}~\bibnamefont{Hartley}},
  \bibinfo{journal}{Phys. Rev. Lett.} \textbf{\bibinfo{volume}{93}},
  \bibinfo{pages}{190402} (\bibinfo{year}{2004}).

\bibitem[{\citenamefont{Wei et~al.}(2006)\citenamefont{Wei, Liu, Sun, and
  Nori}}]{Wei2006}
\bibinfo{author}{\bibfnamefont{L.~F.} \bibnamefont{Wei}},
  \bibinfo{author}{\bibfnamefont{Y.-X.} \bibnamefont{Liu}},
  \bibinfo{author}{\bibfnamefont{C.~P.} \bibnamefont{Sun}}, \bibnamefont{and}
  \bibinfo{author}{\bibfnamefont{F.}~\bibnamefont{Nori}},
  \bibinfo{journal}{Phys. Rev. Lett.} \textbf{\bibinfo{volume}{97}},
  \bibinfo{pages}{237201} (\bibinfo{year}{2006}).

\bibitem[{\citenamefont{Xue et~al.}(2007{\natexlab{a}})\citenamefont{Xue,
  Zhong, Li, and Sun}}]{Xue2007prb}
\bibinfo{author}{\bibfnamefont{F.}~\bibnamefont{Xue}},
  \bibinfo{author}{\bibfnamefont{L.}~\bibnamefont{Zhong}},
  \bibinfo{author}{\bibfnamefont{Y.}~\bibnamefont{Li}}, \bibnamefont{and}
  \bibinfo{author}{\bibfnamefont{C.~P.} \bibnamefont{Sun}},
  \bibinfo{journal}{Phys. Rev. B} \textbf{\bibinfo{volume}{75}},
  \bibinfo{pages}{033407} (\bibinfo{year}{2007}{\natexlab{a}}).

\bibitem[{\citenamefont{Xue et~al.}(2007{\natexlab{b}})\citenamefont{Xue, Wang,
  Sun, Okamoto, Yamaguchi, and Semba}}]{Xue2007NJP}
\bibinfo{author}{\bibfnamefont{F.}~\bibnamefont{Xue}},
  \bibinfo{author}{\bibfnamefont{Y.~D.} \bibnamefont{Wang}},
  \bibinfo{author}{\bibfnamefont{C.~P.} \bibnamefont{Sun}},
  \bibinfo{author}{\bibfnamefont{H.}~\bibnamefont{Okamoto}},
  \bibinfo{author}{\bibfnamefont{H.}~\bibnamefont{Yamaguchi}},
  \bibnamefont{and} \bibinfo{author}{\bibfnamefont{K.}~\bibnamefont{Semba}},
  \bibinfo{journal}{New J. of Phys.} \textbf{\bibinfo{volume}{9}},
  \bibinfo{pages}{35} (\bibinfo{year}{2007}{\natexlab{b}}).

\bibitem[{\citenamefont{Hu and Nori}(1996)}]{Hu1996a}
\bibinfo{author}{\bibfnamefont{X.}~\bibnamefont{Hu}} \bibnamefont{and}
  \bibinfo{author}{\bibfnamefont{F.}~\bibnamefont{Nori}},
  \bibinfo{journal}{Phys. Rev. Lett.} \textbf{\bibinfo{volume}{76}},
  \bibinfo{pages}{2294} (\bibinfo{year}{1996}).

\bibitem{Hu1996b}
\bibinfo{author}{\bibfnamefont{X.}~\bibnamefont{Hu}} \bibnamefont{and}
  \bibinfo{author}{\bibfnamefont{F.}~\bibnamefont{Nori}},
  \bibinfo{journal}{Phys. Rev. B} \textbf{\bibinfo{volume}{53}},
  \bibinfo{pages}{2419} (\bibinfo{year}{1996}{\natexlab{b}}).

\bibitem[{\citenamefont{Hu and Nori}(1997)}]{Hu1997}
\bibinfo{author}{\bibfnamefont{X.}~\bibnamefont{Hu}} \bibnamefont{and}
  \bibinfo{author}{\bibfnamefont{F.}~\bibnamefont{Nori}},
  \bibinfo{journal}{Phys. Rev. Lett.} \textbf{\bibinfo{volume}{79}},
  \bibinfo{pages}{4605} (\bibinfo{year}{1997}).

\bibitem{Hu1999}
\bibinfo{author}{\bibfnamefont{X.}~\bibnamefont{Hu}} \bibnamefont{and}
  \bibinfo{author}{\bibfnamefont{F.}~\bibnamefont{Nori}},
  \bibinfo{journal}{Physica B} \textbf{\bibinfo{volume}{263}},
  \bibinfo{pages}{16} (\bibinfo{year}{1999}).

\bibitem[{\citenamefont{Savel'ev and Nori}(2004)}]{Savel'ev2004Dec}
\bibinfo{author}{\bibfnamefont{S.}~\bibnamefont{Savel'ev}} \bibnamefont{and}
  \bibinfo{author}{\bibfnamefont{F.}~\bibnamefont{Nori}},
  \bibinfo{journal}{Phys. Rev. B} \textbf{\bibinfo{volume}{70}},
  \bibinfo{pages}{214415} (\bibinfo{year}{2004}).

\bibitem[{\citenamefont{Savel'ev
  et~al.}(2006{\natexlab{a}})\citenamefont{Savel'ev, Hu, and
  Nori}}]{Savel'ev2006NJP}
\bibinfo{author}{\bibfnamefont{S.}~\bibnamefont{Savel'ev}},
  \bibinfo{author}{\bibfnamefont{X.}~\bibnamefont{Hu}}, \bibnamefont{and}
  \bibinfo{author}{\bibfnamefont{F.}~\bibnamefont{Nori}}, \bibinfo{journal}{New
  J. of Phys.} \textbf{\bibinfo{volume}{8}}, \bibinfo{pages}{105}
  (\bibinfo{year}{2006}{\natexlab{a}}); cond-mat/0601019.

\bibitem[{\citenamefont{Sergey et~al.}(2007)\citenamefont{Sergey, Rakhmanov,
  Xuedong, Kasumov, and Franco}}]{Savel'ev2007}
\bibinfo{author}{\bibfnamefont{S.}~\bibnamefont{Savel'ev}},
  \bibinfo{author}{\bibfnamefont{A.~L.} \bibnamefont{Rakhmanov}},
  \bibinfo{author}{\bibfnamefont{X.}~\bibnamefont{Hu}},
  \bibinfo{author}{\bibfnamefont{A.}~\bibnamefont{Kasumov}}, \bibnamefont{and}
  \bibinfo{author}{\bibfnamefont{F.}~\bibnamefont{Nori}},
  \bibinfo{journal}{Phys. Rev. B} \textbf{\bibinfo{volume}{75}},
  \bibinfo{pages}{165417} (\bibinfo{year}{2007}).

\bibitem{Chan2001}
\bibinfo{author}{\bibfnamefont{H.~B.} \bibnamefont{Chan}},
  \bibinfo{author}{\bibfnamefont{V.~A}~\bibnamefont{Aksyuk}},
  \bibinfo{author}{\bibfnamefont{R.~N}~\bibnamefont{Kleiman}},
  \bibinfo{author}{\bibfnamefont{D.~J}~\bibnamefont{Bishop}}, \bibnamefont{and}
  \bibinfo{author}{\bibfnamefont{F.}~\bibnamefont{Capasso}},
  \bibinfo{journal}{Phys. Rev. Lett.}
  \textbf{\bibinfo{volume}{87}}, \bibinfo{pages}{211801}
  (\bibinfo{year}{2001}).

\bibitem[{\citenamefont{Munday et~al.}(2005)\citenamefont{Munday, Iannuzzi,
  Barash, and Capasso}}]{Munday2005}
\bibinfo{author}{\bibfnamefont{J.~N.} \bibnamefont{Munday}},
  \bibinfo{author}{\bibfnamefont{D.}~\bibnamefont{Iannuzzi}},
  \bibinfo{author}{\bibfnamefont{Y.}~\bibnamefont{Barash}}, \bibnamefont{and}
  \bibinfo{author}{\bibfnamefont{F.}~\bibnamefont{Capasso}},
  \bibinfo{journal}{Phys. Rev. A}
  \textbf{\bibinfo{volume}{71}}, \bibinfo{pages}{042102}
  (\bibinfo{year}{2005}).

\bibitem[{\citenamefont{Munday et~al.}(2006)\citenamefont{Munday, Iannuzzi, and
  Capasso}}]{Munday2006}
\bibinfo{author}{\bibfnamefont{J.~N.} \bibnamefont{Munday}},
  \bibinfo{author}{\bibfnamefont{D.}~\bibnamefont{Iannuzzi}}, \bibnamefont{and}
  \bibinfo{author}{\bibfnamefont{F.}~\bibnamefont{Capasso}},
  \bibinfo{journal}{New J. of Phys.} \textbf{\bibinfo{volume}{8}},
  \bibinfo{pages}{244} (\bibinfo{year}{2006}).

\bibitem[{\citenamefont{Capasso et~al.}(2007)\citenamefont{Capasso, Munday,
  Iannuzzi, and Chan}}]{Capasso2007}
\bibinfo{author}{\bibfnamefont{F.}~\bibnamefont{Capasso}},
  \bibinfo{author}{\bibfnamefont{J.~N.} \bibnamefont{Munday}},
  \bibinfo{author}{\bibfnamefont{D.}~\bibnamefont{Iannuzzi}}, \bibnamefont{and}
  \bibinfo{author}{\bibfnamefont{H.~B.} \bibnamefont{Chan}},
  \bibinfo{journal}{IEEE J. of Selected Topics in Quantum Electronics}
  \textbf{\bibinfo{volume}{13}}, \bibinfo{pages}{400} (\bibinfo{year}{2007}).

\bibitem[{\citenamefont{Leggett}(2002)}]{Leggett2002}
\bibinfo{author}{\bibfnamefont{A.~J.} \bibnamefont{Leggett}},
  \bibinfo{journal}{J. of Phys.: Cond. Matt.}
  \textbf{\bibinfo{volume}{14}}, \bibinfo{pages}{R415} (\bibinfo{year}{2002}).

\bibitem[{\citenamefont{Metzger and Karrai}(2004)}]{Metzger2004}
\bibinfo{author}{\bibfnamefont{C.~H.} \bibnamefont{Metzger}} \bibnamefont{and}
  \bibinfo{author}{\bibfnamefont{K.}~\bibnamefont{Karrai}},
  \bibinfo{journal}{Nature} \textbf{\bibinfo{volume}{432}},
  \bibinfo{pages}{1002} (\bibinfo{year}{2004}).

\bibitem[{\citenamefont{Arcizet et~al.}(2006)\citenamefont{Arcizet, Cohadon,
  Briant, Pinard, and Heidmann}}]{Arcizet2006}
\bibinfo{author}{\bibfnamefont{O.}~\bibnamefont{Arcizet}},
  \bibinfo{author}{\bibfnamefont{R.~F.} \bibnamefont{Cohadon}},
  \bibinfo{author}{\bibfnamefont{T.}~\bibnamefont{Briant}},
  \bibinfo{author}{\bibfnamefont{M.}~\bibnamefont{Pinard}}, \bibnamefont{and}
  \bibinfo{author}{\bibfnamefont{A.}~\bibnamefont{Heidmann}},
  \bibinfo{journal}{Nature (London)} \textbf{\bibinfo{volume}{444}},
  \bibinfo{pages}{71} (\bibinfo{year}{2006}).

\bibitem[{\citenamefont{Gigan et~al.}(2006)\citenamefont{Gigan, Bohm,
  Paternostro, Blaser, Langer, Hertzberg, Schwab, Bauerle, Aspelmeyer, and
  Zeilinger}}]{Gigan2006}
\bibinfo{author}{\bibfnamefont{S.}~\bibnamefont{Gigan}},
  \bibinfo{author}{\bibfnamefont{H.~R.} \bibnamefont{Bohm}},
  \bibinfo{author}{\bibfnamefont{M.}~\bibnamefont{Paternostro}},
  \bibinfo{author}{\bibfnamefont{F.}~\bibnamefont{Blaser}},
  \bibinfo{author}{\bibfnamefont{G.}~\bibnamefont{Langer}},
  \bibinfo{author}{\bibfnamefont{J.~B.} \bibnamefont{Hertzberg}},
  \bibinfo{author}{\bibfnamefont{K.~C.} \bibnamefont{Schwab}},
  \bibinfo{author}{\bibfnamefont{D.}~\bibnamefont{Bauerle}},
  \bibinfo{author}{\bibfnamefont{M.}~\bibnamefont{Aspelmeyer}},
  \bibnamefont{and}
  \bibinfo{author}{\bibfnamefont{A.}~\bibnamefont{Zeilinger}},
  \bibinfo{journal}{Nature (London)} \textbf{\bibinfo{volume}{444}},
  \bibinfo{pages}{67} (\bibinfo{year}{2006}).

\bibitem[{\citenamefont{Kleckner and Bouwmeester}(2006)}]{Kleckner2006}
\bibinfo{author}{\bibfnamefont{D.}~\bibnamefont{Kleckner}} \bibnamefont{and}
  \bibinfo{author}{\bibfnamefont{D.}~\bibnamefont{Bouwmeester}},
  \bibinfo{journal}{Nature (London)} \textbf{\bibinfo{volume}{444}},
  \bibinfo{pages}{75} (\bibinfo{year}{2006}).

\bibitem[{\citenamefont{Schliesser et~al.}(2006)\citenamefont{Schliesser,
  Del'Haye, Nooshi, Vahala, and Kippenberg}}]{Schliesser2006}
\bibinfo{author}{\bibfnamefont{A.}~\bibnamefont{Schliesser}},
  \bibinfo{author}{\bibfnamefont{P.}~\bibnamefont{Del'Haye}},
  \bibinfo{author}{\bibfnamefont{N.}~\bibnamefont{Nooshi}},
  \bibinfo{author}{\bibfnamefont{K.~J.} \bibnamefont{Vahala}},
  \bibnamefont{and} \bibinfo{author}{\bibfnamefont{T.~J.}
  \bibnamefont{Kippenberg}}, \bibinfo{journal}{Phys. Rev. Lett.}
  \textbf{\bibinfo{volume}{97}}, \bibinfo{pages}{243905}
  (\bibinfo{year}{2006}).

\bibitem[{\citenamefont{Poggio et~al.}(2007)\citenamefont{Poggio, Degen, Mamin,
  and Rugar}}]{Poggio2007}
\bibinfo{author}{\bibfnamefont{M.}~\bibnamefont{Poggio}},
  \bibinfo{author}{\bibfnamefont{C.~L.} \bibnamefont{Degen}},
  \bibinfo{author}{\bibfnamefont{H.~J.} \bibnamefont{Mamin}}, \bibnamefont{and}
  \bibinfo{author}{\bibfnamefont{D.}~\bibnamefont{Rugar}},
\bibinfo{journal}{Phys. Rev. Lett.}
  \textbf{\bibinfo{volume}{99}}, \bibinfo{pages}{017201}
  (\bibinfo{year}{2007}).

\bibitem[{\citenamefont{Martin et~al.}(2004)\citenamefont{Martin, Shnirman,
  Tian, and Zoller}}]{Martin2004Mar}
\bibinfo{author}{\bibfnamefont{I.}~\bibnamefont{Martin}},
  \bibinfo{author}{\bibfnamefont{A.}~\bibnamefont{Shnirman}},
  \bibinfo{author}{\bibfnamefont{L.}~\bibnamefont{Tian}}, \bibnamefont{and}
  \bibinfo{author}{\bibfnamefont{P.}~\bibnamefont{Zoller}},
  \bibinfo{journal}{Phys. Rev. B} \textbf{\bibinfo{volume}{69}},
  \bibinfo{pages}{125339} (\bibinfo{year}{2004}).

\bibitem[{\citenamefont{Zhang et~al.}(2005)\citenamefont{Zhang, Wang, and
  Sun}}]{Zhang2005}
\bibinfo{author}{\bibfnamefont{P.}~\bibnamefont{Zhang}},
  \bibinfo{author}{\bibfnamefont{Y.~D.} \bibnamefont{Wang}}, \bibnamefont{and}
  \bibinfo{author}{\bibfnamefont{C.~P.} \bibnamefont{Sun}},
  \bibinfo{journal}{Phys. Rev. Lett.} \textbf{\bibinfo{volume}{95}},
  \bibinfo{pages}{097204} (\bibinfo{year}{2005}).


\bibitem{Hensinger2005}
\bibinfo{author}{\bibfnamefont{W.~K.}~\bibnamefont{Hensinger}},
  \bibinfo{author}{\bibfnamefont{D.~W.} \bibnamefont{Utami}},
  \bibinfo{author}{\bibfnamefont{H.-S.} \bibnamefont{Goan}},
  \bibinfo{author}{\bibfnamefont{K.-C.} \bibnamefont{Schwab}},
  \bibinfo{author}{\bibfnamefont{C.} \bibnamefont{Monroe}},
  \bibnamefont{and}
  \bibinfo{author}{\bibfnamefont{G.~J.} \bibnamefont{Milburn}},
  \bibinfo{journal}{Phys. Rev. A} \textbf{\bibinfo{volume}{72}},
  \bibinfo{pages}{041405(R)} (\bibinfo{year}{2005}).


\bibitem[{\citenamefont{Wineland et~al.}(2006)\citenamefont{Wineland, Britton,
  Epstein, Leibfried, Blakestad, Brown, Jost, Langer, Ozeri, Seidelin
  et~al.}}]{Wineland2006}
\bibinfo{author}{\bibfnamefont{D.~J.} \bibnamefont{Wineland}},
  \bibinfo{author}{\bibfnamefont{J.}~\bibnamefont{Britton}},
  \bibinfo{author}{\bibfnamefont{R.~J.} \bibnamefont{Epstein}},
  \bibinfo{author}{\bibfnamefont{D.}~\bibnamefont{Leibfried}},
  \bibinfo{author}{\bibfnamefont{R.~B.} \bibnamefont{Blakestad}},
  \bibinfo{author}{\bibfnamefont{K.}~\bibnamefont{Brown}},
  \bibinfo{author}{\bibfnamefont{J.~D.} \bibnamefont{Jost}},
  \bibinfo{author}{\bibfnamefont{C.}~\bibnamefont{Langer}},
  \bibinfo{author}{\bibfnamefont{R.}~\bibnamefont{Ozeri}},
  \bibinfo{author}{\bibfnamefont{S.}~\bibnamefont{Seidelin}}, \bibnamefont{and}
  \bibinfo{author}{\bibfnamefont{J.}~\bibnamefont{Wesenberg}},
\bibinfo{journal}{quant-ph/0606180}.

\bibitem[{\citenamefont{Naik et~al.}(2006)\citenamefont{Naik, Buu, LaHaye,
  Armour, Clerk, Blencowe, and Schwab}}]{Naik2006}
\bibinfo{author}{\bibfnamefont{A.}~\bibnamefont{Naik}},
  \bibinfo{author}{\bibfnamefont{O.}~\bibnamefont{Buu}},
  \bibinfo{author}{\bibfnamefont{M.~D.} \bibnamefont{LaHaye}},
  \bibinfo{author}{\bibfnamefont{A.~D.} \bibnamefont{Armour}},
  \bibinfo{author}{\bibfnamefont{A.~A.} \bibnamefont{Clerk}},
  \bibinfo{author}{\bibfnamefont{M.~P.} \bibnamefont{Blencowe}},
  \bibnamefont{and} \bibinfo{author}{\bibfnamefont{K.~C.}
  \bibnamefont{Schwab}}, \bibinfo{journal}{Nature}
  \textbf{\bibinfo{volume}{443}}, \bibinfo{pages}{193} (\bibinfo{year}{2006}).

\bibitem{Wang2007}
\bibinfo{author}{\bibfnamefont{Y.~D.} \bibnamefont{Wang}},
\bibinfo{author}{\bibfnamefont{K.} \bibnamefont{Semba}},
\bibnamefont{and}
\bibinfo{author}{\bibfnamefont{H.} \bibnamefont{Yamaguchi}},
  \bibinfo{journal}{arXiv: 0704.2462 [cond-mat.mes-hall]}.

\bibitem{Zhao2007}
\bibinfo{author}{\bibfnamefont{N.} \bibnamefont{Zhao}},
\bibinfo{author}{\bibfnamefont{D.L.} \bibnamefont{Zhou}},
\bibinfo{author}{\bibfnamefont{J.-L.} \bibnamefont{Zhu}},
\bibnamefont{and}
\bibinfo{author}{\bibfnamefont{C. P.} \bibnamefont{Sun}},
  \bibinfo{journal}{arXiv: 0705.1964 [cond-mat.mes-hall]}.

\bibitem[{\citenamefont{Bernad et~al.}(2006)\citenamefont{Bernad, Diosi, and
  Geszti}}]{Bernad2006}
\bibinfo{author}{\bibfnamefont{J.~Z.} \bibnamefont{Bernad}},
  \bibinfo{author}{\bibfnamefont{L.}~\bibnamefont{Diosi}}, \bibnamefont{and}
  \bibinfo{author}{\bibfnamefont{T.}~\bibnamefont{Geszti}},
  \bibinfo{journal}{Phys. Rev. Lett.} \textbf{\bibinfo{volume}{97}},
  \bibinfo{pages}{250404} (\bibinfo{year}{2006}).

\bibitem[{\citenamefont{You and Nori}(2003)}]{You2003prb}
\bibinfo{author}{\bibfnamefont{J.~Q.} \bibnamefont{You}} \bibnamefont{and}
  \bibinfo{author}{\bibfnamefont{F.}~\bibnamefont{Nori}},
  \bibinfo{journal}{Phys. Rev. B} \textbf{\bibinfo{volume}{68}},
  \bibinfo{pages}{064509} (\bibinfo{year}{2003}).

\bibitem[{\citenamefont{You and Nori}(2005)}]{You2005PT}
\bibinfo{author}{\bibfnamefont{J.~Q.} \bibnamefont{You}} \bibnamefont{and}
  \bibinfo{author}{\bibfnamefont{F.}~\bibnamefont{Nori}},
  \bibinfo{journal}{Physics Today} \textbf{\bibinfo{volume}{58}}(11),
  \bibinfo{pages}{42} (\bibinfo{year}{2005}).

\bibitem[{\citenamefont{Wallraff et~al.}(2004)\citenamefont{Wallraff, Schuster,
  Blais, Frunzio, Huang, Majer, Kumar, Girvin, and Schoelkopf}}]{Wallraff2004}
\bibinfo{author}{\bibfnamefont{A.}~\bibnamefont{Wallraff}},
  \bibinfo{author}{\bibfnamefont{D.~I.} \bibnamefont{Schuster}},
  \bibinfo{author}{\bibfnamefont{A.}~\bibnamefont{Blais}},
  \bibinfo{author}{\bibfnamefont{L.}~\bibnamefont{Frunzio}},
  \bibinfo{author}{\bibfnamefont{R.~S.} \bibnamefont{Huang}},
  \bibinfo{author}{\bibfnamefont{J.}~\bibnamefont{Majer}},
  \bibinfo{author}{\bibfnamefont{S.}~\bibnamefont{Kumar}},
  \bibinfo{author}{\bibfnamefont{S.~M.} \bibnamefont{Girvin}},
  \bibnamefont{and} \bibinfo{author}{\bibfnamefont{R.~J.}
  \bibnamefont{Schoelkopf}}, \bibinfo{journal}{Nature}
  \textbf{\bibinfo{volume}{431}}, \bibinfo{pages}{162} (\bibinfo{year}{2004}).

\bibitem[{\citenamefont{Schuster et~al.}(2007)\citenamefont{Schuster, Houck,
  Schreier, Wallraff, Gambetta, Blais, Frunzio, Majer, Johnson, Devoret
  et~al.}}]{Schuster2007}
\bibinfo{author}{\bibfnamefont{D.~I.} \bibnamefont{Schuster}},
  \bibinfo{author}{\bibfnamefont{A.~A.} \bibnamefont{Houck}},
  \bibinfo{author}{\bibfnamefont{J.~A.} \bibnamefont{Schreier}},
  \bibinfo{author}{\bibfnamefont{A.}~\bibnamefont{Wallraff}},
  \bibinfo{author}{\bibfnamefont{J.~M.} \bibnamefont{Gambetta}},
  \bibinfo{author}{\bibfnamefont{A.}~\bibnamefont{Blais}},
  \bibinfo{author}{\bibfnamefont{L.}~\bibnamefont{Frunzio}},
  \bibinfo{author}{\bibfnamefont{J.}~\bibnamefont{Majer}},
  \bibinfo{author}{\bibfnamefont{B.}~\bibnamefont{Johnson}},
  \bibinfo{author}{\bibfnamefont{M.~H.} \bibnamefont{Devoret}},
  \bibinfo{author}{\bibfnamefont{S.~M.}~\bibnamefont{Girvin}},
  \bibnamefont{and}
  \bibinfo{author}{\bibfnamefont{R.~J.} \bibnamefont{Schoelkopf}},
 \bibinfo{journal}{Nature}
  \textbf{\bibinfo{volume}{445}}, \bibinfo{pages}{515} (\bibinfo{year}{2007}).

\bibitem{Milonni2004}
\bibinfo{author}{\bibfnamefont{P.~W.} \bibnamefont{Milonni}},
  \bibnamefont{and}
  \bibinfo{author}{\bibfnamefont{B.~M.} \bibnamefont{Chernobrod}},
 \bibinfo{journal}{Nature}
  \textbf{\bibinfo{volume}{432}}, \bibinfo{pages}{965} (\bibinfo{year}{2004}).

\bibitem{Dykman1984}
Soviet Scientific Reviews. Section A: Physics Reviews Vol.5. Edited
by I. M. Khalatnikov (1984).

\bibitem{Brown2007}
\bibinfo{author}{\bibfnamefont{K.~R.}~\bibnamefont{Brown}},
  \bibinfo{author}{\bibfnamefont{J.}~\bibnamefont{Britton}},
  \bibinfo{author}{\bibfnamefont{R.~J.}~\bibnamefont{Epstein}},
  \bibinfo{author}{\bibfnamefont{J.}~\bibnamefont{Chiaverini}},
  \bibinfo{author}{\bibfnamefont{D.}~\bibnamefont{Leibfried}}, \bibnamefont{and}
    \bibinfo{author}{\bibfnamefont{D.~J.}~\bibnamefont{Wineland}},
  \bibinfo{journal}{Phys. Rev. Lett.} \textbf{\bibinfo{volume}{99}},
  \bibinfo{pages}{137205} (\bibinfo{year}{2007}).

\bibitem[{\citenamefont{Sidles et~al.}(1995)\citenamefont{Sidles, L.Garbini,
  and al}}]{Sidles1995}
\bibinfo{author}{\bibfnamefont{J.~A.}~\bibnamefont{Sidles}},
  \bibinfo{author}{\bibfnamefont{J.~L.}~\bibnamefont{Garbini}},
  \bibinfo{author}{\bibfnamefont{K.~J}~\bibnamefont{Bruland}},
  \bibinfo{author}{\bibfnamefont{D.}~\bibnamefont{Rugar}},
  \bibinfo{author}{\bibfnamefont{O.}~\bibnamefont{Z\"uger}},
  \bibinfo{author}{\bibfnamefont{S.}~\bibnamefont{Hoen}}, \bibnamefont{and}
    \bibinfo{author}{\bibfnamefont{C.~S}~\bibnamefont{Yannoni}},
  \bibinfo{journal}{Rev. Mod. Phys.} \textbf{\bibinfo{volume}{67}},
  \bibinfo{pages}{249} (\bibinfo{year}{1995}).

\bibitem{Weissbook2001}
\bibinfo{author}{\bibfnamefont{U.} \bibnamefont{Weiss}},
  \emph{\bibinfo{title}{Quantum dissipative system}} (\bibinfo{publisher}{Second edition, World Scientific},
  \bibinfo{address}{Singapore}, \bibinfo{year}{2001}).

\bibitem[{\citenamefont{Wilson-Rae et~al.}(2007)\citenamefont{Wilson-Rae,
  Nooshi, Zwerger, and Kippenberg}}]{Wilson-Rae2007}
\bibinfo{author}{\bibfnamefont{I.}~\bibnamefont{Wilson-Rae}},
  \bibinfo{author}{\bibfnamefont{N.}~\bibnamefont{Nooshi}},
  \bibinfo{author}{\bibfnamefont{W.}~\bibnamefont{Zwerger}}, \bibnamefont{and}
  \bibinfo{author}{\bibfnamefont{T.~J}~\bibnamefont{Kippenberg}},
  \bibinfo{journal}{Phys. Rev. Lett.} \textbf{\bibinfo{volume}{99}},
  \bibinfo{pages}{093901} (\bibinfo{year}{2007}).

\bibitem[{\citenamefont{Marquardt et~al.}(2007)\citenamefont{Marquardt, Chen,
  Clerk, and Girvin}}]{Marquardt2007}
\bibinfo{author}{\bibfnamefont{F.}~\bibnamefont{Marquardt}},
  \bibinfo{author}{\bibfnamefont{J.~P.} \bibnamefont{Chen}},
  \bibinfo{author}{\bibfnamefont{A.~A.} \bibnamefont{Clerk}}, \bibnamefont{and}
  \bibinfo{author}{\bibfnamefont{S.~M.} \bibnamefont{Girvin}},
  \bibinfo{journal}{Phys. Rev. Lett.} \textbf{\bibinfo{volume}{99}},
  \bibinfo{pages}{093902} (\bibinfo{year}{2007}).

\bibitem{Grajcar2007}
M. Grajar, S. Ashhab, J. R. Johanson, and F. Nori,
arXiv:0709.3775v1.

\end{thebibliography}

\end{document}